\documentclass[conference]{IEEEtran}
\usepackage{blindtext, graphicx}
\usepackage{xcolor}
\usepackage{float}
\usepackage{environ}
\usepackage{datetime}
\usepackage[english]{babel}
\usepackage{caption}
\usepackage{alltt}
\usepackage{array}
\newcolumntype{P}[1]{>{\centering\arraybackslash}p{#1}}
\usepackage{color}

\usepackage[linesnumbered,ruled,vlined]{algorithm2e}
\usepackage{eurosym}
\usepackage{amsmath}
\usepackage{wasysym}
\usepackage{mathtools}
\usepackage{multirow}
\usepackage[space]{grffile}

\ifCLASSINFOpdf
\else
\fi
\begin{document}

\bstctlcite{IEEEexample:BSTcontrol}

%
% paper title
% can use linebreaks \\ within to get better formatting as desired
\title{Hierarchical Aggregation Approach for Distributed clustering of spatial datasets}

%\author{~~~~~}
\author{\IEEEauthorblockN{Malika Bendechache}
\IEEEauthorblockA{Insight Centre for Data Analytics
 \\School of Computer Science\\
University College Dublin\\
Dublin, Ireland\\
Email: malika.bendechache@ucdconnect.ie}
\and
\IEEEauthorblockN{Nhien-An Le-Khac}
\IEEEauthorblockA{School of Computer Science\\ 
University College Dublin\\
Dublin, Ireland\\
Email: an.lekhac@ucd.ie}

\and
\IEEEauthorblockN{M-Tahar Kechadi}
\IEEEauthorblockA{Insight Centre for Data Analytics
\\School of Computer Science\\
University College Dublin\\
Dublin, Ireland\\
Email: tahar.kechadi@ucd.ie}}

% make the title area
\maketitle
\thispagestyle{plain}
\pagestyle{plain}
\raggedbottom % eliminate big spaces between sections

\begin{abstract}
%\boldmath

  In this paper, we present a new approach of distributed clustering for spatial datasets,
  based on an  innovative and efficient aggregation technique.   This distributed approach
  consists of two phases: 1) local clustering phase, where each node performs a clustering
  on its  local data,  2) aggregation phase,  where the local  clusters are  aggregated to
  produce global  clusters.  This  approach is  characterised by the  fact that  the local
  clusters are  represented in a  simple and efficient way.  And The aggregation  phase is
  designed in such  a way that the final clusters are compact  and accurate while the
  overall process is  efficient in both response time and  memory allocation. We evaluated
  the  approach  with  different  datasets   and  compared  it  to  well-known  clustering
  techniques.  The experimental  results  show that  our approach  is  very promising  and
  outperforms all those algorithms.

\end{abstract}

% Note that keywords are not normally used for peerreview papers.

% For peer review papers, you can put extra information on the cover
% page as needed:
% \ifCLASSOPTIONpeerreview
% \begin{center} \bfseries EDICS Category: 3-BBND \end{center}
% \fi
%
% For peerreview papers, this IEEEtran command inserts a page break and
% creates the second title. It will be ignored for other modes.
\IEEEpeerreviewmaketitle

\section{Introduction}

Spatio-temporal   datasets   are    often   very   large   and    difficult   to   analyse
~\cite{dunham-03}. Traditional centralised  data management and mining  techniques are not
adequate, as  they do  not consider  all the  issues of  data-driven applications  such as
scalability  in   both  response  time   and  accuracy  of  solutions,   distribution  and
heterogeneity ~\cite{Bertolotto-07}. In addition, transferring  a huge amount of data over
the  network is  not  an efficient  strategy  and may  not be  possible  for security  and
protection reasons. 

Distributed  data mining  (DDM) techniques  constitute a  better alternative  as they  are
scalable  and can  deal efficiently  with data  heterogeneity.  Many  DDM methods  such as
distributed association  rules and  distributed classification have  been proposed  in the
last decade  and most of them  are based on performing  partial analysis on local  data at
individual sites  followed by the generation  of global models by  aggregating those local
results \cite{Aouad-07,L.Aouad-09,L.Aouad6-10,Le-Khac-10}.   However, only  a few  of them
concern distributed clustering. Among these  many parallel clustering algorithms have been
proposed  ~\cite{Aouad-07s,   Dhillon-99,  Ester-96}.    They  are  classified   into  two
categories.  The  first consists of methods  requiring multiple rounds of  message passing
and significant amount of synchronisations.  The  second category consists of methods that
build  local   clustering  models  and  then   aggregate  them  to  build   global  models
~\cite{Laloux-11}.  Most of  the parallel approaches need  either multiple synchronisation
constraints between processes or a global view of the dataset, or both ~\cite{Le-Khac2-07,L.Aouad3-07}. For  distributed approaches, the  aggregation phase is very  costly, and
therefore, it needs to be optimised.

In this paper, we present an approach that reduces the complexity of the aggregation phase.  It reduces significantly the  amount of information
exchanged during the aggregation phase, and  generates automatically the correct number of
clusters.  In a case  study, it was shown that the data exchanged  is reduced by more than
$98\%$ of the original datasets.

\section{Dynamic Distributed Clustering}
\label{sec:DDC}
\begin{figure}[H]
    \centering
    \begin{center}
    \includegraphics[width=\columnwidth]{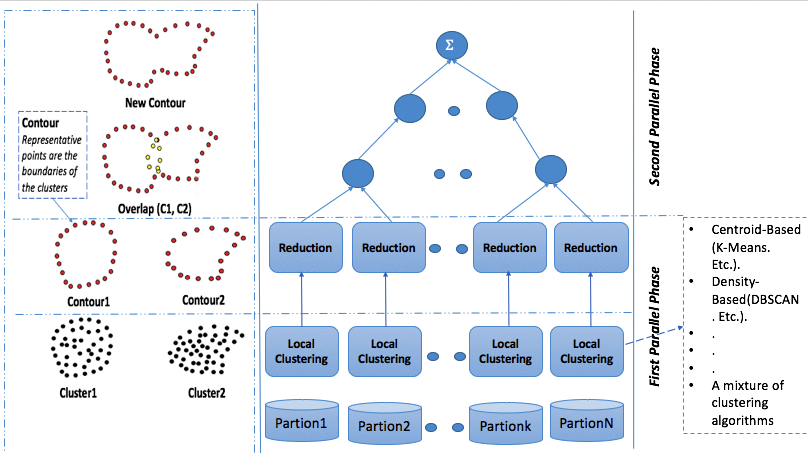}
    \caption{An overview of the DDC Approach.}
    \label{Archi}
     \end{center}
\end{figure}

The DDC  approach includes two main  steps.  In the first  step, as usual, we  cluster the
datasets  located on  each  node and  select  good local  representatives  for each  local
cluster. This phase is executed in  parallel without communications between the nodes.  In
this phase we can reach a super speed-up. The next phase consists of aggregating the local
clusters.  This operation is executed by certain  nodes of the system, called leaders. The
leaders are elected according to some nodes' characteristics such as their capacity, processing
power, connectivity,  etc.  The leaders are  responsible for merging and  regenerating the
data  objects based  on the  local  cluster representatives.  However, the  nodes need  to
communicate in order to send their local clusters  to the leaders. This process is shown in
Figure \ref{Archi}.

\subsection{Local Models}

Each node can use any technique of clustering on its local dataset. This is one of the key
features of this  approach to deal with  data heterogeneity. Each node  choses a technique
that is  more suitable for  its data. This approach  relaxes the data  pre-processing. For
instance we may  not need to deal  with data consistency between the  nodes' datasets. All
needed is the consistency of the  local cluster's representation. Once extracted the local
clusters will be  used as inputs for  the next phase to generate  global clusters. However
exchanging the local clusters through the network will create significant overheads and
slowdown  hugely the  process.  This  is one  of  the major  problems of  the majority  of
distributed clustering  techniques. To improve  the performance  we propose to  exchange a
minimum number of points/objects. Instead of  sending all the clusters datapoints, we only
exchange their representative points, which constitute $1\%$ to $2\%$ of the total size of
the dataset.

The best way  to represent a spatial cluster is  by its shape and density. The  shape of a
cluster  is represented  by  its  boundary points  (called  contour) ~\cite{  Le-Khac2-07,
  Aouad2-07,N-A-10} (see Figure \ref{DBS}).  Many algorithms for extracting the boundaries
from a cluster can be found in the literature ~\cite{A.Ray-97,M.Melkemi-00}.  Recently, we
developed  a  new  algorithm for  detecting  and  extracting  the  boundary of  a  cluster
~\cite{Laloux-11}. The main concepts of this technique are given below. 

\textbf{Neighbourhood:~} Given  a cluster $C\subseteq\Re^n\equiv\{p_1,  p_2,...,p_n\}$ The
\textit{neighbourhood} $N^{C}(p)$ of a point $p$ in  the cluster $C$ is defined as the set
of points ${p_i} \in C$  so that the distance between $p_i$ and $p$  is less than or equal
to $\varepsilon$:
   \begin{equation}
       N^{C}(p) = \{p_i \in C \mid dist(p, p_i) \leq \varepsilon\}
\end{equation}
In order to determine the boundary points of a cluster, we need to introduce the following
concepts: 

\textbf{Displacement vector:~} A displacement vector from $p_i \in N^{C}(p)$ to the point
$p$ is defined as:

\begin{equation}
     \overrightarrow{\rm V} = \sum_{p_i \in N^{C}(p)}(p - p_i)
\end{equation}
This vector points towards the area of the lowest density of the neighbourhood of $p$.
   
\textbf{Balance  vector:~} The  balance vector  relative to  the point  $p$ is  defined as
follows:
\begin{equation}
  \overrightarrow{\rm b}_{p} =  \begin{cases} \frac{1}{\|\overrightarrow{\rm V}_{p} \|}
    \overrightarrow{\rm V}_{p}  & \mbox{~if~}  {\|\overrightarrow{\rm V}_{p} \|}  > 0\\
    \overrightarrow{\rm 0} & \mbox{Otherwise} 
  \end{cases}
\end{equation}

\begin{figure}
  \centering
  \begin{center}
    \includegraphics[width=3cm,height=3cm]{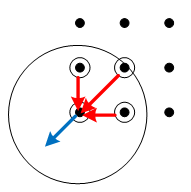}
    \caption{Example of a balance vector.}
    \label{fig1}
  \end{center}
\end{figure}

Note that the balance vector of $p$ points to the least dense area of the neighbourhood of
$p$, the length of  the vector does not hold any  relevant information.  Figure \ref{fig1}
shows an example  of a balance vector.   The neighbourhoods of $p$ are  inside the circle,
the balance vector is represented by the blue colour.

\begin{figure}
  \centering
  \begin{center}
    \includegraphics[width=6cm, height=4cm]{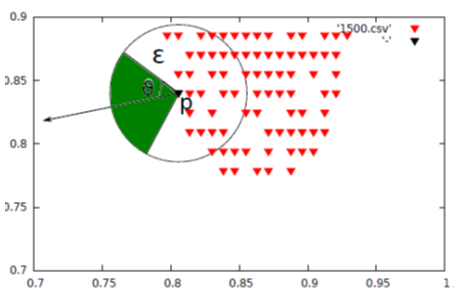}
    \caption{Boundary point check}
    \label{fig2}
  \end{center}
\end{figure}

\textbf{Boundary points:~}  If a  point is  a boundary  point, there  should be  no points
towards the direction of the balance vector.  This property allows us to separate boundary
points and internal points. For each point $p$, it checks for an empty area whose shape is
the intersection  of an hyper-cone of  infinite height, vertex, axis  and aperture $\rho$,
where $\rho$ is a given angle.  As shown in \ref{fig2}, the area checked is highlighted in
green. Formally, a boundary point is described as a Boolean predicate.

%...................fig2 
% ....................................................................

\begin{equation}\label{eq:5}
\small
Boundary(p) = 
\begin{cases} true & \mbox{~if~} (\forall q \in N_\varepsilon^C (p), (q - p)\overrightarrow{\rm
    b}_{b} < \cos(v) \\ 
  false  & \mbox{Otherwise}
\end{cases}
\end{equation}

We can define the boundary $B_C$ of a cluster $C$ as the set of all boundary points in $C$:

\begin{equation}
  B_C = \{p \in C: \mbox{Boundary(p) is true}\}
\end{equation}
   
The algorithm for selecting the boundary  points is described in Algorithm \ref{boundary},
and its complexity is $\mathcal{O}(n\log{n})$.

\begin{algorithm}[!h]
  \SetKwInOut{Input}{input}
  \SetKwInOut{Output}{output}
  \Input {Cluster $C$, set of balance vectors ${\overrightarrow{\rm b}_{p_i}}$, parameter $\nu$.} 
  \Output {Boundary points $B_C$ of cluster $C$}
  
  $B_C \gets C$\;
  \For{every point $p \in B_C$}
  {
    \For{every point $q \in N^C(p)$}
    {
      \If{$(q-p)\overrightarrow{\rm b}\ge \cos(v)$}
      {
        Discard $p$ from $B_C$\;        
        Break\;
      }
    }
  }
  \Return{$B_C$}\;
  \caption{Boundary Detection Algorithm}
  \label{boundary}
\end{algorithm}

Let $C_i$  be a set of  clusters of node  $i$ and $B_{c_j}$  be the boundary of  a cluster
$c_j\in C_i$. The local model $L_i$ is defined by:

\begin{equation}
  L_i = {\bigcup_{j=1}^n	{B_{c_j} }\cup P_i} 
 \end{equation}

Where $P_i$ is the set of internal representatives. 

\subsection{Global Models}

The global clusters are generated during the second phase. This phase consists of two main
steps: 1) each leader collects the local  clusters of its neighbours, 2) the leaders merge
the local  clusters using  the overlay  technique.  The process  of merging  clusters will
continue until we reach the root node. The root node will contain the global clusters (see
Figure \ref{Archi}). During this phase we only exchange the boundaries of the clusters.

The merging process consists of two  steps: boundary merging and regeneration. The merging
is performed by a boundary-based method. Let $L_i$ be a local model received from the site
$i$ and $B_i$ be the set of all boundaries in $L_i$. The global model $G$ is defined by:

\begin{equation}
      G = {\Gamma(\cup B_i), B_i \in L_i}
   \end{equation}

where $\Gamma$ is a merging function. 

\section{DDC Evaluation and Validation}
\label{sec:DDC-EV}
DDC  is  evaluated  using  different  clustering  algorithms.  In  this  paper  we  use  a
centroid-based algorithm (K-Means) and a density-based Algorithm (DBSCAN).
  
\subsection{DDC-K-Means}
\label{DDC-K}
The DDC  with K-Means
(DDC-K-Means) is characterised  by the fact that  in the first phase,  called the parallel
phase, each node $N_i$  of the system executes the K-Means algorithm  on its local dataset
to produce $L_i$ local  clusters and calculate their contours. The rest  of the process is
the same as described above.  
It was shown in \cite{M-Bendechache-15} that DDC-K-Means dynamically determines the number
of the clusters without a priori knowledge about  the data or an estimation process of the
number of the clusters.  DDC-K-Means was compared to two well-known clustering algorithms:
BIRCH ~\cite{Tian-96}  and CURE ~\cite{Sudipto-01}.  The results showed that  it generates
much better  clusters.  Also,  as expected, this  approach runs much  faster than  the two
other algorithms; BIRCH and CURE.

DDC-K-Means does  not need  the number of  global clusters in  advance.  It  is calculated
dynamically. Moreover, each  local clustering $L_i$ needs $K_i$. Let  $\tilde{K_i}$ be the
exact number of local clusters in the node $N_i$,  all it is required is to set $K_i$ such
that $K_i > \tilde{K_i}$. This is much simpler than giving $K_i$, especially when there is
not enough knowledge  about the local dataset characteristics. Nevertheless,  it is indeed
better  to set  $K_i$  as  close as  possible  to $\tilde{K_i}$  in  order  to reduce  the
processing time in calculating the contours and also merging procedure.

However, DDC-K-Means  fails to find good  clusters for datasets with  Non-Covex shapes and
also for  datasets with noises,  this is due to  the fact
that the K-Means  algorithm tends to work with  convex shape only, because it  is based on
the  centroid principle  to generate  clusters.   Moreover, we  can also  notice that  the
results of DDC-K-Means are  even worse with dataset which contains a  big amount of noises
($T_3$, and  $T_4$). In  fact it returns  the whole  dataset with the  noise as  one final
cluster for each dataset  (see Figure \ref{FIG1}).  This is because  K-Means does not deal
with noise.

\subsection{DDC with DBSCAN}

DBSCAN  (Density-Based spatial  Clustering of  Applications  with Noise)  is a  well-known
density based clustering  algorithm capable of discovering clusters  with arbitrary shapes
and eliminating noisy  data ~\cite{M.Ester-96}.  
 
 \subsubsection{\textbf{DBSCAN Complexity}}
\label{DBCom}
DBSCAN visits each point of the dataset, possibly multiple times, depending on the size of
the neighbourhood.  If it performs a neighbourhood operations in $\mathcal{O}(\log n)$, an
overall average complexity  of $\mathcal{O}(n \log n)$ is obtained  if the parameter $Eps$
is  chosen  in  a meaningful  way.   The  worst  case  execution time  complexity  remains
$\mathcal{O}  (n^2)$.   The  distance  matrix  of  size  $\mathcal{O}((n^2-n/2))$  can  be
materialised  to avoid  distance  re-computations, but  this  needs $\mathcal{O}(n^2)$  of
memory, whereas  a non-matrix based implementation  of DBSCAN only needs  $O(n)$ of memory
space.

\subsubsection{\textbf{DDC-DBSCAN Algorithm}}

The approach remains the same; instead of  using K-Means for processing local clusters, we
use DBSCAN. Each node ($n_i$) executes DBSCAN  on its local dataset to produce $K_i$ local
clusters. Once  all the local  clusters are determined,  we calculate their  contours. The
second  phase  is   the  same  and  the   pseudo  code  of  the  algorithm   is  given  in
Algorithm~\ref{clustering}.

\begin{algorithm}[!htb]
  \SetKwInOut{Input}{input}\SetKwInOut{Output}{output}
  \Input {$X_i$: Dataset Fragment, $Eps_i$: Distance $Eps_i$ for $Node_i$, $MinPts_i$: 
    minimum points contain clusters generated by  $Node_i$, $D$: tree degree, $L_i$: Local
    clusters generated by $Node_i$} 
  \Output {$K_g$: Global Clusters (global results)}
  \BlankLine
  \BlankLine
  $level = tree height$\;
  
  \begin{enumerate}
  \item DBSCAN($X_i$. $Eps_i$, $MinPts_i$)\;  \tcp{$Node_i$ executes DBSCAN locally.}
  \item Contour($L_i$)\; \tcp{$Node_i$ executes the Contour algorithm to generate the boundary 
        of each local cluster.} 
  \item $Node_i$ joins a group $G$ of $D$ elements\; \tcp{$Node_i$ joins its neighbourhood} 
  \item Compare cluster of $Node_i$  to other node's clusters in the same group\;
    \tcp{look for overlapping between clusters.} 
  \item j = ElectLeaderNode()\;
    \tcp{Elect a node which will merge the overlapping clusters.} 
  \end{enumerate}
  \If {$i<>j$} {
    Send (contour i, j)\;
  }
  \Else 
  {
    \If {$level>0$}{
      $level$ - - \;
      Repeat 3, 4, and 5 until level=0\;
    }
    \Else
    {
      \Return ($K_g$: $Node_i$' clusters)\;
    }}
  \caption{DDC with DBSCAN.}
  \label{clustering}
\end{algorithm}

\begin{figure}[H]
  \centering
  \begin{center}
    \includegraphics[width=\columnwidth]{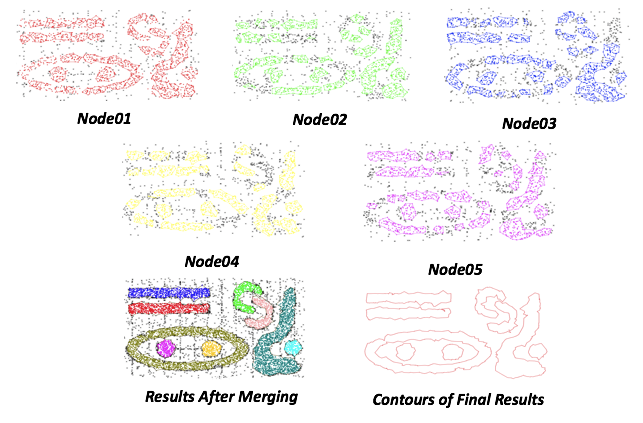}
    \caption{Example of DDC-DBSCAN execution.}
    \label{DBS}
  \end{center}
\end{figure}

Figure  \ref{DBS} illustrates  an  example  of DDC-DBSCAN.   Assume  that the  distributed
computing platform contains five Nodes ($N=5$).   Each Node executes DBSCAN algorithm with
its local  parameters ($Eps_i$, $MinPts_i$)  on its  local dataset. As  it can be  seen in
Figure \ref{DBS} the new approach returned exactly  the right number of clusters and their
shapes. The approach is insensitive to the way the original data was distributed among the
nodes. It  is also insensitive to  noise and outliers. As  we can see, although  each node
executed DBSCAN locally with different parameters.  The global final clusters were correct
even on the noisy dataset ($T_3$) (See Figure \ref{DBS}).

\section{Experimental Results}
\label{sec:ExRes}
In this section, we  study the performance of the DDC-DBSCAN  approach and demonstrate its
effectiveness compared to BIRCH, CURE and  DDC-K-Means. We choose these algorithms because
either  they  are in  the  same  category such  as  BIRCH  which belongs  to  hierarchical
clustering category, or have an efficient optimisation approach, such as CURE.

\textbf{BIRCH}: We used the BIRCH implementation provided in \cite{Tian-96}. It performs a
pre-clustering  and then  uses a centroid-based  hierarchical clustering  algorithm. Note
that the time and  space complexity of this approach is quadratic to  the number of points
after  pre-clustering.  We  set  its parameters  to  the default  values  as suggested  in
\cite{Tian-96}.

\textbf{CURE}:  We used  the implementation  of CURE  provided in  \cite{Sudipto-01}.  The
algorithm uses  representative points  with shrinking  towards the  mean. As  described in
\cite{Sudipto-01},  when  two  clusters  are  merged   in  each  step  of  the  algorithm,
representative points  for the new merged  cluster are selected  from the ones of  the two
original clusters rather than all the points in the merged clusters.

\subsection{Experiments}
We run experiments with different datasets.  We used four types of datasets ($T_1$, $T_2$,
$T_3$  and $T_4$)  with different  shapes and  sizes. These  datasets are  very well-known
benchmarks to  evaluate density-based  clustering algorithms.  All  the four  datasets are
summarised in Table  \ref{table1}.  The number of  points and clusters in  each dataset is
also given. These four datasets contain a set  of shapes or patterns which are not easy to
extract with traditional techniques.

\begin{table}[htb]
  \caption{The datasets used to test the algorithms.}
  \centering 
  \begin{center}
    \begin{tabular}{|c|c|c|c|c|}
      \hline
      \textbf{Type}  & \textbf{Dataset} & \textbf{Description} & \textbf{\#Points} & \textbf{\#Clusters} \\ \hline
      \multirow{1}{*}{\textbf{Without Noise}}  & T1   & \begin{tabular}[c]{@{}c@{}}Round and Oval shape\end{tabular}      & 700   & 5  \\\hline
      \multirow{3}{*}{\textbf{\begin{tabular}[c]{@{}c@{}}With Noise\end{tabular}}} & T2 & \begin{tabular}[c]{@{}c@{}}Different shapes\\ including noise\end{tabular} & 321    & 6     \\ \cline{2-5}  & T3   & \begin{tabular}[c]{@{}c@{}}Different shapes, with \\ some clusters surrounded\\  by others\end{tabular} & 10,000   & 9 \\ \cline{2-5}  & T4 & Different shapes with Noises  & 8,000    & 6    \\ \hline
    \end{tabular}
    \label{table1}
  \end{center}
\end{table}

\subsection{Quality of Clustering}
\label{QC}
\begin{figure*}[!ht]
\centering
\begin{tabular}{c | c | c | c}
  BIRCH & CURE & DDC-K-Means & DDC-DBSCAN \\ \hline \\

  \includegraphics[width=0.18\textwidth, height=0.15\textwidth]{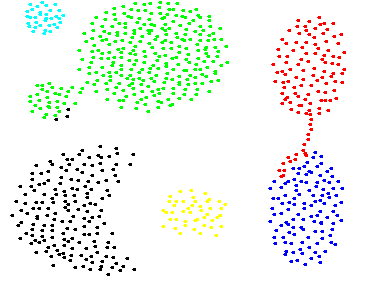} &
    \includegraphics[width=0.18\textwidth, height=0.15\textwidth]{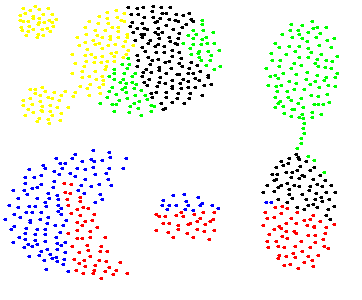} &
    \includegraphics[width=0.18\textwidth, height=0.15\textwidth]{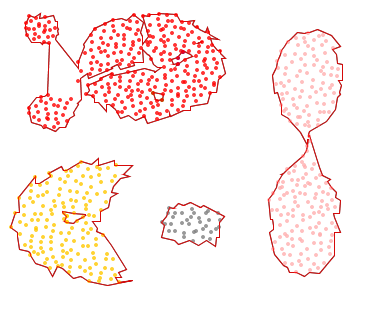} &
    \includegraphics[width=0.18\textwidth, height=0.15\textwidth]{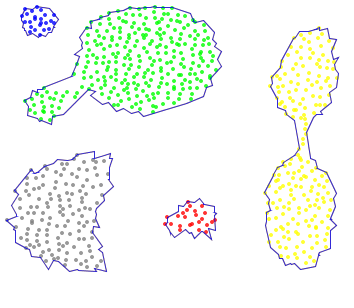} 
 \\ 
    
 \multicolumn{4}{c}{T1}
 \\ 
 
 \includegraphics[width=0.18\textwidth, height=0.15\textwidth]{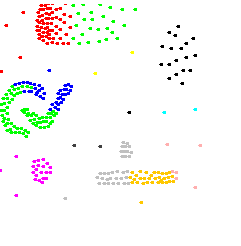} &
    \includegraphics[width=0.18\textwidth, height=0.15\textwidth]{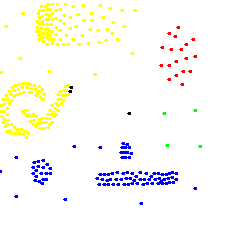} &
    \includegraphics[width=0.18\textwidth, height=0.15\textwidth]{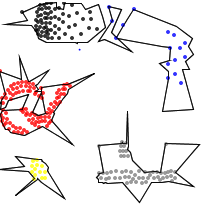} &  \includegraphics[width=0.18\textwidth, height=0.15\textwidth]{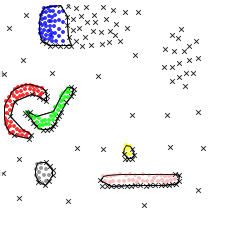}
 \\ 
  \multicolumn{4}{c}{T2}
 
 \\ 
 \includegraphics[width=0.18\textwidth, height=0.15\textwidth]{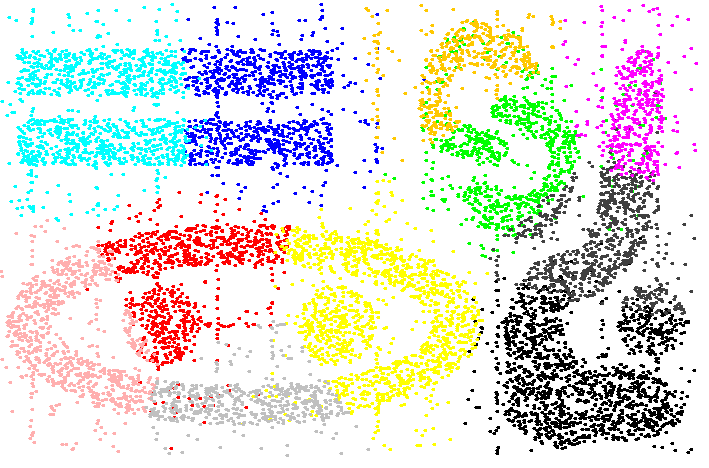} &
    \includegraphics[width=0.18\textwidth, height=0.15\textwidth]{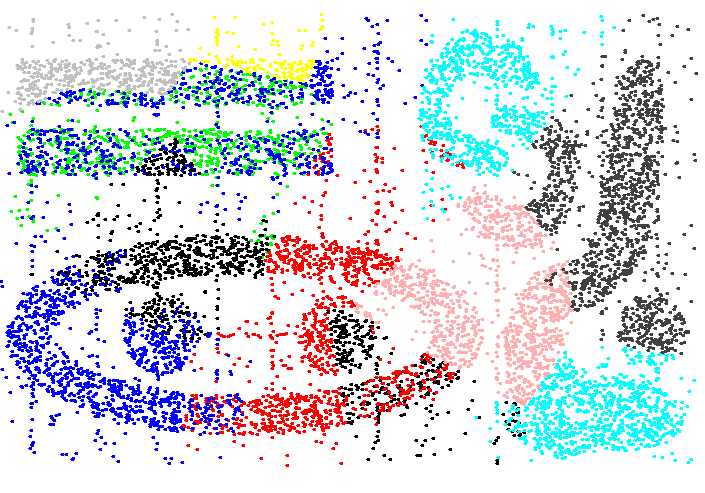} &
    \includegraphics[width=0.18\textwidth, height=0.15\textwidth]{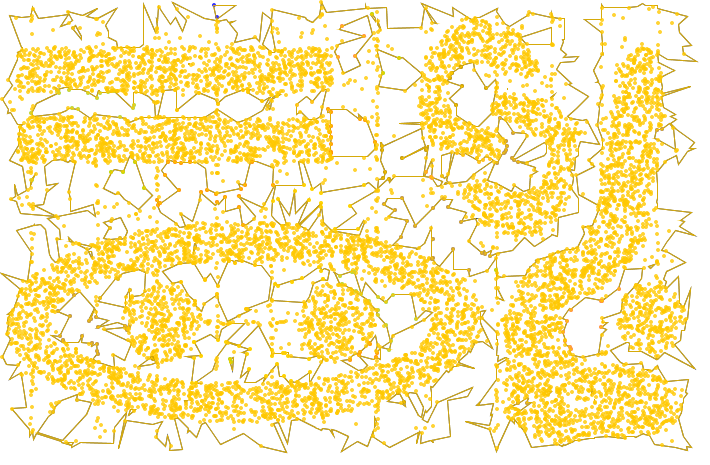} &
     \includegraphics[width=0.18\textwidth, height=0.15\textwidth]{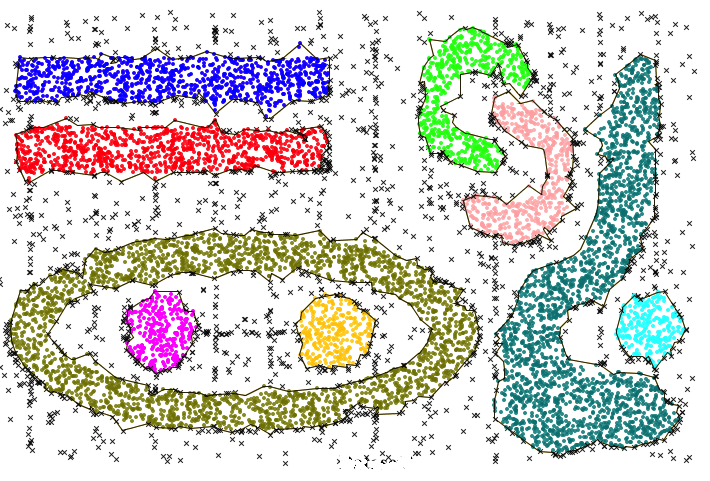}
 \\ 
    
  \multicolumn{4}{c}{T3}
 \\ 
    \includegraphics[width=0.18\textwidth, height=0.15\textwidth]{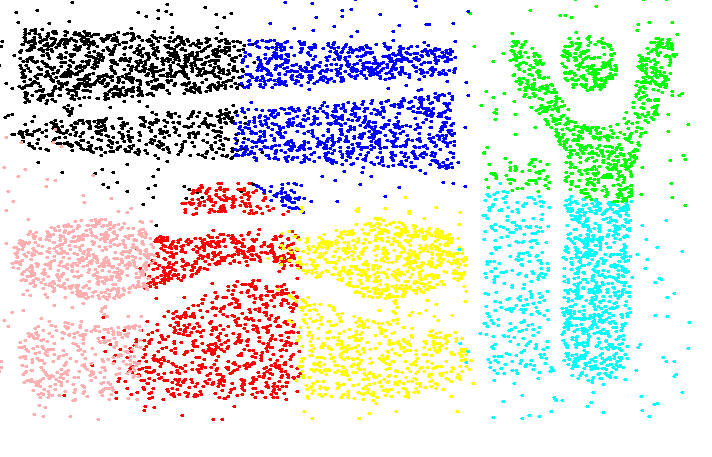} &
    \includegraphics[width=0.18\textwidth, height=0.15\textwidth]{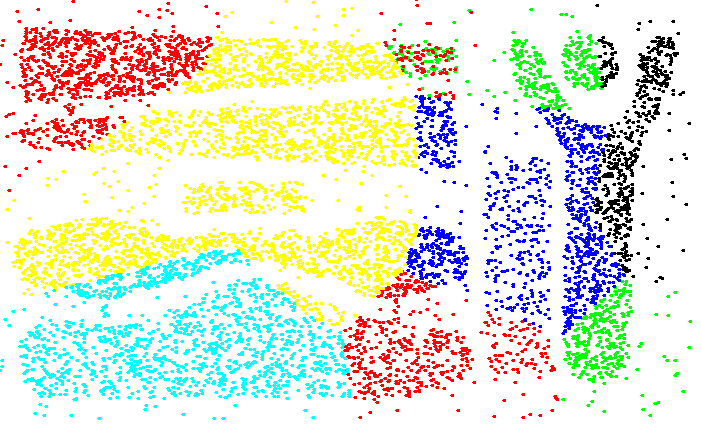} &
    \includegraphics[width=0.18\textwidth, height=0.15\textwidth]{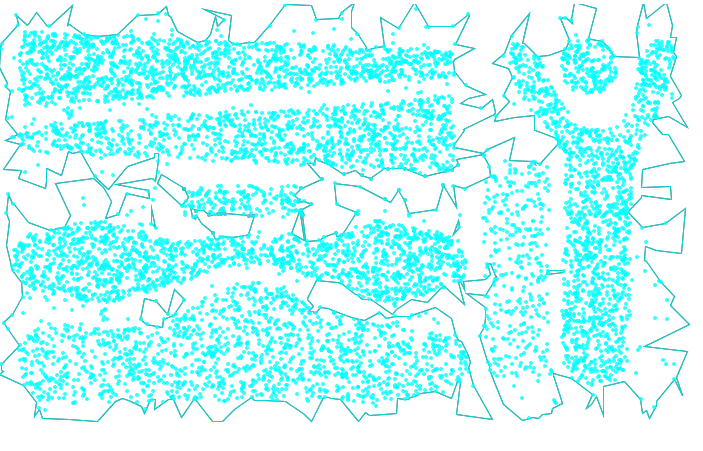} &
    \includegraphics[width=0.18\textwidth, height=0.15\textwidth]{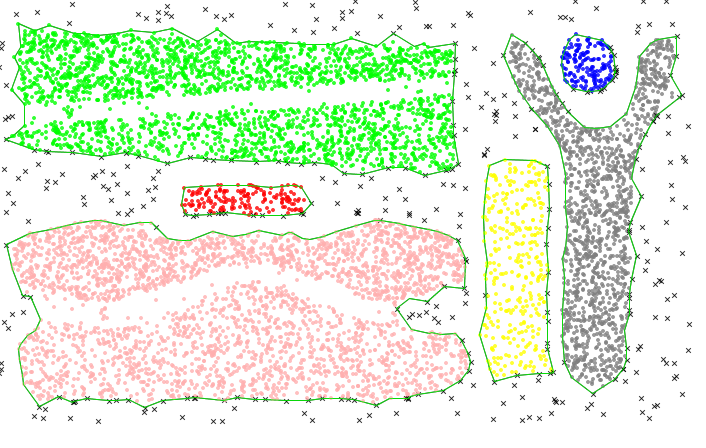} 
 \\ 
 
  \multicolumn{4}{c}{T4}
 \\ 
\end{tabular}
\caption{Comparing the clusters generated across different datasets}
\label{FIG1}
\end{figure*}

We run the four algorithms on the Four  datasets in order to evaluate the quality of their
final  clusters.  Figure \ref{FIG1}  shows  the  clusters returned  by  each  of the  four
algorithms for the datasets without noise ($T_1$) and datasets with noise ($T_2, T_3,$ and
$T_4$). We use different colours to show the clusters returned by each algorithm.

%....................................Convex and Non convex shape................
As expected, BIRCH could not find correct  clusters; it tends to work slightly better with
datasets without noise ($T_1$), as BIRCH does not deal with noise. The results of CURE are
worse and it is not able to extract clusters with non-convex shapes.  We can also see that
CURE does not  deal with noise.  DDC-K-Means  fails to find the correct  final results. In
fact it returns  the whole original dataset  as one final cluster for  each dataset ($T_3$
and  $T_4$) (Which  contain significant  amount  of noise).   This confirms  that the  DDC
technique is sensitive  to the type of  the algorithm chosen for the  first phase. Because
the  second phase  deals only  with the  merging of  the local  clusters whether  they are
correct or  not. This  issue is  corrected by  the DDC-DBSCAN,  as it  is well  suited for
spatial datasets with  or without noise. For  datasets with noise it  eliminates the noise
and outliers. In fact, it generates good  final clusters in datasets that have significant
amount of noise.

As a final observation, these results prove  that the DDC framework is very efficient with
regard to  the accuracy of  its results.  The  only issue is  to choose a  good clustering
algorithm for the  first phase. This can  be done by exploring the  initial datasets along
with  the  question  to  be  answered  and  choose  an  appropriate  clustering  algorithm
accordingly.

Moreover, as for the DDC-K-Means, DDC-DBSCAN is dynamic (the correct number of clusters is
returned  automatically) and  efficient (the  approach  is distributed  and minimises  the
communications).

\subsection{Speed-up}

The goal here  is to study the execution  time of the four algorithms  and demonstrate the
impact of using a parallel and distributed  architecture to deal with the limited capacity
of a centralised system.

As mentioned in  Section \ref{DBCom}, the execution time for  the DDC-DBSCAN algorithm can
be generated in two cases. The first case  is to include the time required to generate the
distance matrix calculation.  The  second case is to suppose that  the distance matrix has
already been generated. The reason for this is that the distance matrix is calculated only
once.

\begin{table}[htb]
\caption{The execution times (ms) of BIRCH,  CURE, DDC-K-Means and DDC-DBSCAN with (w) and
  without (w/o) distance matrix computation.} 
\centering
\begin{tabular}{cr|r|r|r|r|r|}
\cline{3-7}
\multicolumn{2}{l}{} & \multicolumn{5}{|c|}{\textit{\textbf{Execution Time (ms)}}} \\ \cline{2-7} 
\multicolumn{1}{l|}{} & \multicolumn{1}{c|}{{\color[HTML]{000000} }} & \multicolumn{1}{c|}{} & \multicolumn{1}{c|}{}  & \multicolumn{1}{c|}{} & \multicolumn{2}{c|}{\textbf{DDC-DBSCAN}} \\ \cline{6-7} 
\multicolumn{1}{l|}{\multirow{-2}{*}{\textbf{}}} & \multicolumn{1}{c|}{\multirow{-2}{*}{{\color[HTML]{000000} \textbf{SIZE}}}} & \multicolumn{1}{c|}{\multirow{-2}{*}{\textbf{BIRCH}}} & \multicolumn{1}{c|}{\multirow{-2}{*}{\textbf{CURE}}} & \multicolumn{1}{c|}{\multirow{-2}{*}{\textbf{DDC-K-Means}}} & \multicolumn{1}{c|}{\textbf{W}} & \multicolumn{1}{c|}{\textbf{W/O}} \\ \hline
\multicolumn{1}{|c|}{\textbf{T1}} & 700 & 150 & 70045 & 120& 500 & 302 \\ \hline
\multicolumn{1}{|c|}{\textbf{T2}} & 321 & 72 & 47045 & 143 & 504 & 312 \\ \hline
\multicolumn{1}{|c|}{\textbf{T3}} & 10000 & 250 & 141864 & 270 & 836 & 470 \\ \hline
\multicolumn{1}{|c|}{\textbf{T4}} & 8000 & 218 & 92440 & 257 & 608 & 304 \\ \hline
\end{tabular}
\label{Table2}
\end{table}

Table \ref{Table2}  illustrates the execution  times of  the four techniques  on different
datasets. Note that the execution times do  not include the time for post-processing since
these are the same for the four algorithms.

As  mentioned in  Section  \ref{DBCom}, Table  \ref{Table2} confirmed  the  fact that  the
distance  matrix  calculation  in  DBSCAN is  very  significant.   Moreover,  DDC-DBSCAN's
execution time is  much lower than CURE’s  execution times across the  four datasets. Table
\ref{Table2} shows  also that  the DDC-K-Means  is very quick  which is  in line  with its
polynomial computational complexity. BIRCH is also  very fast, however, the quality of its
results  are not  good, it  failed in  finding  the correct  clusters across  all the  four
datasets.

The DDC-DBSCAN is a  bit slower that DDC-K-Means, but it returns  high quality results for
all the tested benchmarks, much better that DDC-K-Means, which has reasonably good results
for convex cluster shapes and very bad results for non-convex cluster shapes.  The overall
results confirm that  the DDC-DBSCAN clustering techniques compares favourably  to all the
tested  algorithms for  the  combined performance  measures (quality  of  the results  and
response time).

%\subsection{Scalability}

%The goal  here is to determine  the effects of  the number of  nodes in the system  on the
%execution times.  The  dataset contains $50,000$ data points. Figure  \ref{SCAL} shows the
%execution time against the number of nodes ($x\_axis$ is in $log_2$) in the system. As one
%can see,  DDC-DBSCAN took only  few seconds (including  the matrix computation's  time) to
%cluster $50,000$ data points in a distributed  system that contains up to $100$ nodes, the
%algorithm took  even less time  when we exclude the  matrix computation's time.  Thus, the
%algorithm can comfortably handle high-dimensional data because of its low complexity.

%\begin{figure}[H]
%\begin{center}
%\includegraphics[width=\columnwidth, height=5.5cm]{SCALUP}
%\caption{Scalability Experiments.}
%\label{SCAL}
%\end{center}
%\end{figure}

%%%%%%%%%%%%%%%%%%%%%%%%%%%AN adding: complexity
\subsection{Computation Complexity}

Let $n$ be  the number of data objects  in the dataset. The complexity of  our approach is
the sum of  the complexities of its  three components: local mining,  local reduction, and
global aggregation.

\paragraph{Phase1: Local clustering}
Assume that the local  clustering algorithm is DBSCAN for all the nodes.  The cost of this
phase is given by:
\[
T_{Phase_1} = 
 \operatorname*{Max}_{i=1}^N(DBSCAN_i) + \operatorname*{Max}_{i=1}^N(Reduction_i)  
\]

Where $N$ is the number of nodes in the  system. The complexity of DBSCAN in the best case
scenario is $\mathcal{O}(n \log  n)$ if the parameter $Eps$ is chosen  in a meaningful way
and also  if we exclude  the distance matrix computation.  Finally, the complexity  of the
local reduction algorithm is $\mathcal{O}(n\log{}n)$.
 
\paragraph{Phase2: Aggregation}
Our  global aggregation  depends  on the  hierarchical combination  of  contours of  local
clusters. As the combination is based on  the intersection of edges from the contours, the
complexity of  this phase  is $\mathcal{O}(v\log{}v  + p)$.   Where $v$  is the  number of
vertices  and $p$  is the  number  of intersections  between edges  of different  contours
(polygons).

\paragraph{Total    complexity}    The   total    complexity    of    our   approach    is
$T_{Total}  = \mathcal{O}(n  \log  n)+  \mathcal{O}(n\log{}n) +  \mathcal{O}(v\log{}v+p)$,
which is:

\[T_{Total} \simeq \mathcal{O}(n \log  n)
\]

%%%%%%%%%%%%%%%%%%%%%%%%%%%End of complexity

\section{Conclusion}
\label{sec:Con}
In this paper, we proposed an efficient and flexible distributed clustering framework that
can work with existing  data mining algorithms.  The framework has  been tested on spatial
datasets using the K-Means and DBSCAN algorithms.  The proposed approach is dynamic, which
solves one  of the  major shortcomings  of K-Means  or DBSCAN.   We proposed  an efficient
aggregation phase,  which reduces considerably the  data exchange between the  leaders and
the system nodes. The size of the data exchange is reduced by about $98\%$.

The DDC approach was tested using various benchmarks. The benchmarks were chosen in such a
way to reflect all the difficulties of clusters extraction. These difficulties include the
shapes of  the clusters (convex  and non-convex), the  data volume, and  the computational
complexity. Experimental results  showed that the approach is very  efficient and can deal
with various situations (various shapes, densities, size, etc.).

As future work, we will try to extend  the framework to non-spatial datasets. We will also
look at the problem of the data and communications reduction during phase two.

% if have a single appendix:
%\appendix[Proof of the Zonklar Equations]
% or
%\appendix  % for no appendix heading
% do not use \section anymore after \appendix, only \section*
% is possibly needed

% use appendices with more than one appendix
% then use \section to start each appendix
% you must declare a \section before using any
% \subsection or using \label (\appendices by itself
% starts a section numbered zero.)
%

\section*{Acknowledgment}
\label{sec:Ack}
The  research work  is  conducted in  the  Insight  Centre for  Data  Analytics, which  is
supported by Science Foundation Ireland under Grant Number SFI/12/RC/2289.
\ifCLASSOPTIONcaptionsoff
  \newpage
\fi

% trigger a \newpage just before the given reference
% number - used to balance the columns on the last page
% adjust value as needed - may need to be readjusted if
% the document is modified later
%\IEEEtriggeratref{8}
% The "triggered" command can be changed if desired:
%\IEEEtriggercmd{\enlargethispage{-5in}}

% references section

% can use a bibliography generated by BibTeX as a .bbl file
% BibTeX documentation can be easily obtained at:
% http://www.ctan.org/tex-archive/biblio/bibtex/contrib/doc/
% The IEEEtran BibTeX style support page is at:
% http://www.michaelshell.org/tex/ieeetran/bibtex/
%\bibliographystyle{IEEEtran}
% argument is your BibTeX string definitions and bibliography database(s)
%\bibliography{IEEEabrv,../bib/paper}
%
% <OR> manually copy in the resultant .bbl file
% set second argument of \begin to the number of references
% (used to reserve space for the reference number labels box)
\bibliography{Mybib}
\bibliographystyle{IEEEtran}
% biography section
% 
% If you have an EPS/PDF photo (graphicx package needed) extra braces are
% needed around the contents of the optional argument to biography to prevent
% the LaTeX parser from getting confused when it sees the complicated
% \includegraphics command within an optional argument. (You could create
% your own custom macro containing the \includegraphics command to make things
% simpler here.)
%\begin{biography}[{\includegraphics[width=1in,height=1.25in,clip,keepaspectratio]{mshell}}]{Michael Shell}
% or if you just want to reserve a space for a photo:

\begin{IEEEbiography}[{\includegraphics[width=1in,height=1.25in,clip,keepaspectratio]{picture}}]{John Doe}

\end{IEEEbiography}

% You can push biographies down or up by placing
% a \vfill before or after them. The appropriate
% use of \vfill depends on what kind of text is
% on the last page and whether or not the columns
% are being equalized.

%\vfill

% Can be used to pull up biographies so that the bottom of the last one
% is flush with the other column.
%\enlargethispage{-5in}

% that's all folks
\end{document}